\documentclass[12pt,a4paper]{article}
\usepackage[utf8]{inputenc}
\usepackage[T2B]{fontenc}
\usepackage[english]{babel}
\usepackage{amsmath}
\usepackage{amsfonts}
\usepackage{amssymb}
\usepackage{graphicx}
\usepackage[left=2.50cm, right=2.50cm, top=2.50cm, bottom=2.50cm]{geometry}
\author{A.I. Maimistov}

\begin{document}
	\begin{center}
		%[2\baselineskip]
		%    \vfill
		\textbf{\LARGE Electromagnetic wave propagation in a nonlinear hyperbolic medium }\\
		
		\bigskip
		
		\textit{ A.I. Maimistov }\\[1\baselineskip]
		%\bigskip
		Department of Solid State Physics and Nanostructures, National Nuclear Research University,
		Moscow Engineering Physics Institute, Moscow 115409, Russia \\
		%\bigskip
		E-mails:~~aimaimistov@gmail.com~
		%\end{center}
		
	\end{center}
	%	\tableofcontents
	\date{date}
	
	\textbf{Abstract}
	
The propagation of a quasi-harmonic electromagnetic wave in a bulk hyperbolic dielectric metamaterial is considered. If the  group velocities dispersion is not taken into account, then wave propagation can be described either by the hyperbolic nonlinear Schrodinger equation or by the hyperbolic Manakov equations. It is shown that the region in the space of wave vectors in which the modulation instability of a spatially homogeneous wave is possible is not limited, in contrast to the case of ordinary media.
	
	\section{Introduction}

The creation and study of new materials, in addition to applied purposes, can become a source of new ideas and theoretical models, as well as modifications of previously known ones. An example of such a situation is the study of composite artificial media --- metamaterials with unusual electrodynamics properties \cite{SSS:01,5,6,8,Ramak:05,Sar:Shal:13}. 
A striking property of such media is negative refraction and the possibility of propagation of backward waves. Optical anisotropic media have recently been added to the number of metamaterials based on nanostructured metal-dielectric media --- hyperbolic materials \cite{Elser:06,Nogin:09,Drachev:13,Pod:Kivsh:13}. For uni-axial anisotropic materials, the refractive index is $n(\theta)$ depending on the angle between the optical axis and the wave vector $\theta$ is determined by the well-known formula
$$
	\frac{\cos^2\theta }{\varepsilon_o} +\frac{\sin^2\theta }{\varepsilon_e} = \frac{1}{n^2(\theta)} .
$$
where $\varepsilon_e$  and  $\varepsilon_o$  are the principal values of the permittivity tensor. In ordinary dielectrics, both of these values are positive, but in a hyperbolic medium $\varepsilon_e$ and $\varepsilon_o$ have different signs. 
Most of the investigations of the optical properties of hyperbolic media are devoted to the study of phenomena near the interface of media: the Purcell effect \cite{Pod:Belo:12,Pod:Belo:13}, the refraction, the reflection and the surface plasmons propagation \cite{Zhao:2013}. Directional waves in the waveguides have been studied in some detail  \cite{Ishii:14,Babicheva:15,Boardm:17}.  Unlike conventional planar waveguides, a finite number of the waveguide modes can exist in hyperbolic waveguides, and each mode has two cutoff frequencies  \cite{Lyash:Maim:15,Lyash:Maim:16,Lyash:Maim:17}. The propagation of electromagnetic waves in the bulk hyperbolic media has been studied to a lesser extent (see reviews \cite{8,Drachev:13}).
 
 In this paper, the  electromagnetic wave propagation in a nonlinear hyperbolic dielectric is considered under the assumption of weak nonlinearity and dispersion.
 
 In a conventional nonlinear medium under these conditions, the scalar wave equation reduces to the nonlinear Schr\"{o}dinger equation (NLS). In the one-dimensional case, we talk about (1+1)NLS, in the two-dimensional case, they talk about (2+1)NLS, where the first digit denotes the spatial dimension. It will be shown here that if we neglect the group velocities dispersion, then in the two-dimensional case instead of (2+1)NLS turns out to be a hyperbolic NLS, which can be defined as (1+2)NLS \footnote{The article \cite{Tekin:21} talks about the (2+1) dimensional de Sitter space-time.}. Since the electromagnetic field has more than one component in an anisotropic medium, and a hyperbolic dielectric is an anisotropic uniaxial material, the system of hyperbolic NLS is needed to describe wave propagation. In particular it is the hyperbolic system of Manakov equations. As a simple illustration of the hyperbolicity of a nonlinear medium, the modulation instability will be considered here.

	\section{Electromagnetic waves in a linear hyperbolic medium}

It is assumed that the axis of the Cartesian coordinate system $X$ is directed along the optical axis of a homogeneous anisotropic uniaxial medium. The other two axes are directed so that the permittivity tensor has a diagonal form in this coordinate system: $\hat{\varepsilon}=\mathrm{diag}(\varepsilon_e,~\varepsilon_o,~\varepsilon_o)$. Let the electromagnetic wave propagate in the direction of the vector $\mathbf{n}$ lying in the plane $XZ$ . In this case, the offset along the $Y$ axis turns out to be a symmetry transformation, as a result of which the fields will not depend on the $y$ variable. In a linear homogeneous isotropic medium, the Maxwell's equations decompose into two systems of equations that describe transverse electric (TE) and transverse magnetic (TM) waves \cite{Tamir:78}. In general, in an anisotropic medium, due to electrical induction, TE and TM waves are connected.
However, if we assume that the coordinate axes (i.e., the orientation of an anisotropic uniaxial medium) are chosen as in the case under consideration, then such a separation is possible. The Maxwell equations for TE and TM waves have the following form
\begin{eqnarray}
	\frac{1}{c}\frac{\partial B_x}{\partial t} =  \frac{\partial E_y}{\partial z}, &~~& \frac{1}{c}\frac{\partial D_x}{\partial t} = \frac{\partial H_y}{\partial z}, \nonumber\\
	-\frac{1}{c}\frac{\partial B_z}{\partial t} = \frac{\partial E_y}{\partial x}, &~~& \frac{1}{c}\frac{\partial D_z}{\partial t} = \frac{\partial H_y}{\partial x}, \label{eq:HM:MaxE:2} \\
	\frac{1}{c}\frac{\partial D_y}{\partial t} = \frac{\partial H_x}{\partial z}- \frac{\partial H_z}{\partial x}, &~~&  
	\frac{1}{c}\frac{\partial B_y}{\partial t} = \frac{\partial E_z}{\partial x} -\frac{\partial E_x}{\partial z},
	\nonumber 
\end{eqnarray} and
$$ \frac{\partial D_x}{\partial x} +\frac{\partial D_z}{\partial z} =0, \quad   \frac{\partial B_x}{\partial x} +\frac{\partial B_z}{\partial z} =0. $$
It is convenient to use the spectral representation, assuming
	$$
	\mathbf{F}(x,z,t) = \int_{-\infty} ^{\infty}\mathbf{\tilde{F}}(x,z,\omega)e^{-i\omega t}\frac{d\omega}{2\pi}, $$
where $\mathbf{F}$ is used to denote the fields $\mathbf{E}$, $\mathbf{B}$ and the inductions $\mathbf{D}$, $\mathbf{H}$. For the Fourier components, the  Maxwell equations are written in the following form
\begin{eqnarray}
	ik_0 \tilde{B}_x =  -\frac{\partial \tilde{E}_y}{\partial z}, &~~& ik_0 \tilde{D}_x = \frac{\partial \tilde{H}_y}{\partial z}, \nonumber\\
	ik_0 \tilde{B}_z = \frac{\partial \tilde{E}_y}{\partial x},    &~~& ik_0 \tilde{D}_z = -\frac{\partial \tilde{H}_y}{\partial x}, \label{eq:HM:MaxE:3}\\
	ik_0 \tilde{D}_y = \frac{\partial \tilde{H}_z}{\partial x}- \frac{\partial \tilde{H}_x}{\partial z}, &~~&  
	ik_0 \tilde{B}_y = \frac{\partial \tilde{E}_z}{\partial x} -\frac{\partial \tilde{E}_x}{\partial z}.
	\nonumber 
\end{eqnarray} 
	
Let the magnetic permeability be a constant, i.e, the medium is magnetically isotropic. The electrical induction is related to the electric field by the following relations
	\begin{equation}\label{eq:HM:WE:D1E}
	\tilde{D}_x= \varepsilon_e \tilde{E}_x,\quad \tilde{D}_y= \varepsilon_o \tilde{E}_y,\quad \tilde{D}_z= \varepsilon_o \tilde{E}_z.
	\end{equation}
Thus, two systems of equations are obtained, which may be related due to the nonlinear properties of the medium, but in a linear medium they are independent:	
	\begin{eqnarray}
	ik_0 \mu \tilde{H}_x =  -\frac{\partial \tilde{E}_y}{\partial z}, &~~& ik_0 \varepsilon_e \tilde{E}_x = \frac{\partial \tilde{H}_y}{\partial z}, \nonumber\\
	ik_0 \mu \tilde{H}_z = \frac{\partial \tilde{E}_y}{\partial x},    &~~& ik_0 \varepsilon_o \tilde{E}_z = -\frac{\partial H_y}{\partial x}, \label{eq:HM:MaxE:4}\\
	ik_0 \varepsilon_o\tilde{E}_y = \frac{\partial \tilde{H}_z}{\partial x}- \frac{\partial \tilde{H}_x}{\partial z}, &~~&  
	ik_0 \mu \tilde{H}_y = \frac{\partial \tilde{E}_z}{\partial x} -\frac{\partial \tilde{E}_x}{\partial z}.
	\nonumber 
	\end{eqnarray}
The equations in the left column of the system of equations (\ref{eq:HM:MaxE:4}) describe TE waves. The components of the magnetic field are expressed in terms of $E_y$, which satisfies the equation
	\begin{equation}\label{eq:HM:MaxE:TE}
	\frac{\partial^2 \tilde{E}_y}{\partial x^2} + \frac{\partial^2 \tilde{E}_y}{\partial z^2} +k_0^2\mu\varepsilon_o \tilde{E}_y =0.
	\end{equation}
The dispersion equation for TE waves has the form
	$$k_x^2 +k_z^2 = k_0^2\mu\varepsilon_o.$$
That is, the TE wave is an ordinary wave for which the medium looks like isotropic if $\mu>0$ and $\varepsilon_o>0$. If $\varepsilon_o<0$ then waves do not propagate at $\mu>0$ and propagate at $\mu< 0$.
	
The equations in the right column of the system of equations (\ref{eq:HM:MaxE:4}) describe TM waves. The components of the electric field are expressed in terms of $\tilde{H}_y$, which satisfies the equation	\begin{equation}\label{eq:HM:MaxE:TM}
	\frac{1}{\varepsilon_o}\frac{\partial^2 \tilde{H}_y}{\partial x^2} + \frac{1}{\varepsilon_e}\frac{\partial^2 \tilde{H}_y}{\partial z^2} +k_0^2\mu \tilde{H}_y =0.
\end{equation}
The dispersion equation for TM waves has the form
\begin{equation}\label{eq:HM:MaxE:TM:disper}
	\frac{k_x^2}{\varepsilon_o} +\frac{k_z^2}{\varepsilon_e} = k_0^2\mu 
\end{equation}
Hence, the TM wave is an extraordinary one for which the medium looks like anisotropic. If the signs $\varepsilon_o$ and $\varepsilon_e$ are opposite, then such an anisotropic medium will be hyperbolic.

The propagation of an ordinary wave in a nonlinear medium in the case of weak nonlinearity and dispersion is well described by the nonlinear Schr\"{o}dinger equation and its various generalizations. The results of research in this area can be found in
 \cite{Sukhorukov:Vinog:90,Maim:Bashar:book,Akh:Ank:2003,Kivsh:Agr:2005,Rys:Trub:2017}. Less attention was paid to the extraordinary waves \cite{Boardm:17,Boardman:2016}. For this reason, the propagation of extraordinary waves will be considered further.
 
	\section{Propagation of extraordinary waves in a nonlinear medium}
	
	\subsection{Wave equations in the weak dispersion and nonlinearity approximation of }
	
The starting point of the study for the TM wave, taking into account the nonlinear polarization, are the equations
\begin{eqnarray}
	&&  \frac{\partial \tilde{H}_y}{\partial z} =  ik_0 \varepsilon_e \tilde{E}_x  +ik_04\pi \tilde{P}_x = ik_0\tilde{D}_x, \nonumber\\
	&& \frac{\partial \tilde{H}_y}{\partial x} =-ik_0 \varepsilon_o \tilde{E}_z -ik_04\pi \tilde{P}_z =-ik_0\tilde{D}_z, \label{eq:HM:Maxwell:TM:1}\\
	&&  ik_0 \mu \tilde{H}_y =  \frac{\partial \tilde{E}_x}{\partial z} - \frac{\partial \tilde{E}_z}{\partial x}.
	\nonumber 
\end{eqnarray}
where	$\tilde{P}_x$ and $\tilde{P}_z$ are polarizations describing the nonlinear response of the medium. Two equations follow from the last equation of this system
	\begin{eqnarray}
	&& -k_0^2\mu \tilde{D}_x =  \frac{\partial^2 \tilde{E}_x}{\partial z^2} -\frac{\partial^2 \tilde{E}_z}{\partial z \partial x} , \label{eq:HM:Maxwell:TM:2x}\\
	&& -k_0^2\mu \tilde{D}_z =  \frac{\partial^2 \tilde{E}_z}{\partial x^2} -\frac{\partial^2 \tilde{E}_x}{\partial z \partial x}, \label{eq:HM:Maxwell:TM:2z}
\end{eqnarray}
Since there are supposed to be no free charges and currents, then $\mathrm{div}\mathbf{D} =0$, that is,
	$$ \frac{\partial \tilde{D}_x}{\partial x} +\frac{\partial \tilde{D}_z}{\partial z} =0.$$
	
In weakly nonlinear media the nonlinear contributions to polarization are small, so that in the case of a spatially homogeneous medium, the relation
	\begin{equation}\label{eq:HM:Maxwell:TM:3}
	\varepsilon_e\frac{\partial \tilde{E}_x}{\partial x} +\varepsilon_o\frac{\partial \tilde{E}_z}{\partial z} =0.
	\end{equation}  is valid.
Then from (\ref{eq:HM:Maxwell:TM:2x}) follows the wave equation for $E_x$:
	\begin{equation}\label{eq:HM:Maxwell:TM:3x}
	\frac{\partial^2 \tilde{E}_x}{\partial z^2} + \frac{\varepsilon_e}{\varepsilon_o}\frac{\partial^2 \tilde{E}_x}{\partial x^2} +k_0^2\mu \tilde{D}_x =0.
	\end{equation}
Similarly from (\ref{eq:HM:Maxwell:TM:2z}) and (\ref{eq:HM:Maxwell:TM:3}) the wave equation for $\tilde{E}_z$:
	\begin{equation}\label{eq:HM:Maxwell:TM:3z}
	\frac{\partial^2 \tilde{E}_z}{\partial x^2} + \frac{\varepsilon_o}{\varepsilon_e}\frac{\partial^2 \tilde{E}_z}{\partial z^2} +k_0^2\mu \tilde{D}_z =0.
	\end{equation}
Here  $\tilde{D}_x= \varepsilon_e \tilde{E}_x  +4\pi \tilde{P}_x$,  $\tilde{D}_z= \varepsilon_o \tilde{E}_z  +4\pi \tilde{P}_z$. Thus, the system of wave equations for the components of an extraordinary wave in a uniaxial weakly nonlinear medium  has the following form
	\begin{eqnarray}
	&& 	 \frac{1}{\varepsilon_o}\frac{\partial^2 \tilde{E}_x}{\partial x^2} +\frac{1}{\varepsilon_e}\frac{\partial^2 \tilde{E}_x}{\partial z^2} +k_0^2\mu \tilde{E}_x + \frac{k_0^24\pi\mu}{\varepsilon_e}\tilde{P}_x  =0, \label{eq:HM:Maxwell:TM:4x}\\
	&& 	\frac{1}{\varepsilon_e}\frac{\partial^2 \tilde{E}_x}{\partial z^2}+\frac{1}{\varepsilon_e}\frac{\partial^2 \tilde{E}_z}{\partial z^2}  +k_0^2\mu \tilde{E}_z + \frac{k_0^24\pi\mu}{\varepsilon_o}\tilde{P}_z  =0. \label{eq:HM:Maxwell:TM:4z}
	\end{eqnarray}	
These equations are valid both for an ordinary anisotropic medium when $\varepsilon_o>0$ and $\varepsilon_e>0$, and for a hyperbolic medium where $\varepsilon_o>0$ and $\varepsilon_e<0$ or $\varepsilon_o<0$ and $\varepsilon_e>0$.

In addition to the assumption of weak nonlinearity, it will be assumed that the principal values of the permittivity tensor do not depend on frequency, at least in the region near the carrier frequency $\omega_0$, where the spectral functions $\tilde{E}_{x,z}(\omega)$ are nonzero. By performing the inverse Fourier transform to (\ref{eq:HM:Maxwell:TM:4x}) and (\ref{eq:HM:Maxwell:TM:4z}) in these approximations, it is possible to obtain a system of equations for the fields $E_{x,z}(x,z, t)$. If we determine the following values $E_1=E_x$, $E_2=E_z$, $P_1=P_x$ and $P_2=P_z$, then the system of wave equations for the components of the electric field of a TM wave propagating in a hyperbolic nonlinear medium can be written in the following form
\begin{eqnarray}
	&& 	\frac{\partial^2 E_1}{\partial x^2} - \frac{\partial^2 E_1}{\partial z_1^2}  - \frac{n^2}{c^2}\frac{\partial^2 E_1}{\partial t^2}  = -N_1, \label{eq:HM:Maxwell:TM:6x} \\
	&& \frac{\partial^2 E_2}{\partial x^2} - \frac{\partial^2 E_2}{\partial z_1^2} - \frac{n^2}{c^2}\frac{\partial^2 E_2}{\partial t^2}  = -N_2, \label{eq:HM:Maxwell:TM:6z}
\end{eqnarray}
where the new independent variable is used $z_1 = z(\mid \varepsilon_e/\varepsilon_o\mid)^{1/2}$, $n^2=\mu\varepsilon_{o}$ is the square of the refractive index for an ordinary wave\footnote{It was implied here that we are talking about a hyperbolic medium with $\varepsilon_o>0$. Otherwise, put $\mu\varepsilon_{o} =-n^2$. }, and $N_{1,2}$ are the contributions from nonlinear polarization
$$
	N_1 = \frac{4\pi\mu}{c^2}\int_{-\infty}^{\infty} \theta \omega^2 P_1(x,z,\omega)e^{i\omega t}d\omega, \quad 
	N_2 = \frac{4\pi\mu}{c^2}\int_{-\infty}^{\infty}  \omega^2 P_2(x,z,\omega)e^{i\omega t}d\omega,
$$
where $\theta = \varepsilon_o/\varepsilon_e$.
	
It should be noted that in the case of $E_2=0$ and $P_2=0$, the system of equations (\ref{eq:HM:Maxwell:TM:6x}) and (\ref{eq:HM:Maxwell:TM:6z}) is reduced to one equation that was used in a series of papers
\cite{Smol:Narim:10} --\cite{Tekin:21}, in which hyperbolic media were considered from the point of view of simulating the physical processes of quantum field theory and gravity.

	\subsection{Transition to slowly varying envelopes approximation }

In many cases, when the duration of electromagnetic pulses is much longer than the period of the field strength oscillation, the approximation of the slowly varying envelope of the pulse is used \cite{Sukhorukov:Vinog:90,Rys:Trub:2017}. It is assumed that the fields are quasi-harmonic, that is
$$
	E_j (x,z,t) = A_j(x,z,t)e^{-i\omega_0t+ipx+iqz_1} + c.c.,\quad j=1,~2,
$$
where the envelope of the electric field strength $A_j$ varies slowly over time, so that $\mid\partial A/\partial t\mid\ll\omega_0\mid A\mid$. Equations (\ref{eq:HM:Maxwell:TM:6x}) and (\ref{eq:HM:Maxwell:TM:6z}) lead to the following equations for envelopes $A_j$
	$$\left( \frac{\partial^2 A_j}{\partial x^2} - \frac{\partial^2 A_j}{\partial z_1^2}\right)  + $$
	$$+2i \left( p\frac{\partial A_j}{\partial x} -q \frac{\partial A_j}{\partial z_1} + \frac{n^2\omega_0}{c^2}\frac{\partial A_j}{\partial t} \right) + \left( q^2-p^2 + \frac{n^2\omega_0^2}{c^2}\right) A_j=-\mathcal{N}_j,$$
where the transition to slowly varying envelopes of nonlinear polarization is carried out in the usual way:
$$
	\mathcal{N}_1 = \frac{4\pi\mu \omega_0^2}{c^2} \theta \mathcal{P}_1(x,z,t), \quad 
	\mathcal{N}_2 = \frac{4\pi\mu \omega_0^2}{c^2}  \mathcal{P}_2(x,z,t).
$$
	
The expression in the last term on the left side of this equation is zero, since $q^2-p^2=n^2\omega_0^2/c^2$
is the dispersion relation for an extraordinary wave. To do this, it is enough to note that the components of the wave vector are related to the values $p$ and $q$ as follows $k_x=p$ and $k_z=q(\mid\varepsilon_e/\varepsilon_o\mid)^{1/2}$. Using the operator $\hat{L}$
$$
	\hat{L} = 	2i \left( p\frac{\partial }{\partial x} -q \frac{\partial }{\partial z_1} + \frac{n^2\omega_0}{c^2}\frac{\partial }{\partial t} \right) + \left( \frac{\partial^2 }{\partial x^2} - \frac{\partial^2 }{\partial z_1^2}\right), 
$$
the equations for $A_j$ can be written in the following form
	\begin{equation}\label{eq:HM:Maxwell:TM:8}
	\hat{L}A_j = -\mathcal{N}_j.
	\end{equation}
Replacing variables $\zeta =\tau - px-qz_1$, $\eta =qx+pz_1$, $\tau_1=\tau$ allows to convert (\ref{eq:HM:Maxwell:TM:8}), having received	
$$
	i \frac{\partial A_j}{\partial \tau_1} +\frac{1}{2}\left( \frac{\partial^2}{\partial \zeta^2}-\frac{\partial^2}{\partial \eta^2}\right)A_j = -\frac{c^2}{n^2\omega_0^2}\mathcal{N}_j.
$$
Given the definition of $\mathcal{N}_j$, we can write the system of equations for $A_j$ in the following form
\begin{eqnarray}
	&& i \frac{\partial A_1}{\partial \tau_1} +\frac{1}{2}\left( \frac{\partial^2 A_1}{\partial \zeta^2}-\frac{\partial^2 A_1}{\partial \eta^2}\right)=  -\frac{4\pi}{\varepsilon_e} \mathcal{P}_1 ,  \label{eq:HM:Maxwell:TM:10a}\\
	&&   i \frac{\partial A_2}{\partial \tau_1} +\frac{1}{2}\left( \frac{\partial^2 A_2}{\partial \zeta^2}-\frac{\partial^2 A_2}{\partial \eta^2}\right)=  -\frac{4\pi}{\varepsilon_o} \mathcal{P}_2 .\label{eq:HM:Maxwell:TM:10b}
\end{eqnarray}
	
In these equations, expressions for the nonlinear polarizations  were not detailed. Hence (\ref{eq:HM:Maxwell:TM:10a}) and (\ref{eq:HM:Maxwell:TM:10b}) can be used to study waves in the cubical nonlinear media, in the media with competing nonlinear responses, or in the media with saturating nonlinearities.

It should be noted that since the signs $\varepsilon_o$ and $\varepsilon_e$ are opposite for hyperbolic media, the signs before the terms reflecting nonlinear responses for different components of the electric field vector will be opposite.

	\subsection{Some special cases  }

If an electromagnetic wave has only one component, for example $A_1$, then its propagation in the approximation considered here can be described by the single equation
	\begin{equation}\label{eq:HM:Maxwell:TM:11}
	i \frac{\partial A_1}{\partial \tau_1} +\frac{1}{2}\left( \frac{\partial^2 A_1}{\partial \zeta^2}-\frac{\partial^2 A_1}{\partial \eta^2}\right)= - \frac{4\pi}{\varepsilon_e} \mathcal{P}_1 ,
	\end{equation}
If at the same time $\mathcal{P}_1 = \chi^{(3)} \mid A_1\mid^2 A_1$, then we get a hyperbolic NLS equation:	
 \cite{AiLin:2010} -- \cite{Baleanu:Park:2021}:
	\begin{equation}\label{eq:HM:Maxwell:TM:HNLS}
	i \frac{\partial A_1}{\partial \tau_1} +\frac{1}{2}\left( \frac{\partial^2 A_1}{\partial \zeta^2}-\frac{\partial^2 A_1}{\partial \eta^2}\right)=  -\frac{4\pi \chi^{(3)} }{\varepsilon_e} \mid A_1\mid ^2 A_1.
	\end{equation}
Here the sign $\varepsilon_e$ defines the role of the Kerr (cubic) nonlinearity: the either self-focusing or defocusing can occur. For (\ref{eq:HM:Maxwell:TM:HNLS}) exact solutions in the form of solitary waves are known
\cite{AiLin:2010,Durur:2020,TalaTebue:2020,Baleanu:Park:2021}. 
The conservation laws were obtained \cite{Aliuy:2018} in addition to solutions of the type of solitary waves. The modulation instability of a spatially homogeneous wave is considered in \cite{Apeanti:2019}.
Although solitary waves in these works are called solitons, full integrability (\ref{eq:HM:Maxwell:TM:HNLS}), as far as is known, has not been established and the term "soliton" \ is not used quite legally.
	
In the case when the nonlinear properties of the medium are described by polarizations
$$
	\mathcal{P}_1 = \chi^{(3)}\left(  \mid A_1\mid ^2 +\mid A_2\mid ^2\right)  A_1,\quad 
	\mathcal{P}_2 = \chi^{(3)}\left(  \mid A_1\mid ^2 +\mid A_2\mid ^2\right)  A_2,
$$
the system of equations (\ref{eq:HM:Maxwell:TM:10a}) and (\ref{eq:HM:Maxwell:TM:10b}) takes the following form
	\begin{eqnarray}
	&& i \frac{\partial A_1}{\partial \tau_1} +\frac{1}{2}\left( \frac{\partial^2 A_1}{\partial \zeta^2}-\frac{\partial^2 A_1}{\partial \eta^2}\right)=  -\frac{4\pi \chi^{(3)}}{\varepsilon_e} \left(  \mid A_1\mid ^2 +\mid A_2\mid ^2\right)  A_1,  \label{eq:HM:Maxwell:TM:12a}\\
	&&   i \frac{\partial A_2}{\partial \tau_1} +\frac{1}{2}\left( \frac{\partial^2 A_2}{\partial \zeta^2}-\frac{\partial^2 A_2}{\partial \eta^2}\right)=  -\frac{4\pi \chi^{(3)}}{\varepsilon_o} \left(  \mid A_1\mid ^2 +\mid A_2\mid ^2\right)  A_2. \label{eq:HM:Maxwell:TM:12b}
	\end{eqnarray}
	
If the fields did not depend on the variable $\zeta$ (or $\eta$), then the resulting system of equations would coincide with the Manakov \cite{Manakov:73} equations. For this reason, the equations (\ref{eq:HM:Maxwell:TM:12a}) and (\ref{eq:HM:Maxwell:TM:12b}) can be called \emph{ the hyperbolic Manakov system of equations}.	Initially, the Manakov equations were used to describe the self-focusing of a plane beam. It is possible that (\ref{eq:HM:Maxwell:TM:12a}) and (\ref{eq:HM:Maxwell:TM:12b}) can be used to study self-focusing in a nonlinear hyperbolic medium.

	\section{Modulation instability of a homogeneous wave in a hyperbolic medium}
	
It is known that the usual nonlinear Schrodinger equation and the Manakov system of equations have solutions that describe a spatially homogeneous wave \cite{Akh:Ank:2003,Rhys:Trub:2017}. With an increase in the amplitude of this wave, the harmonic perturbations of the homogeneous solution begin to increase exponentially over time.
This phenomenon is called modulation instability (MI). This is one of their typical phenomena in the evolution of nonlinear waves. For hyperbolic NLS (\ref{eq:HM:Maxwell:TM:HNLS}) the MI phenomenon was investigated in \cite{Apeanti:2019}. As an example of the application of the generalization of the Manakov system obtained here to the hyperbolic case, the MI based on the system of equations (\ref{eq:HM:Maxwell:TM:12a}) and (\ref{eq:HM:Maxwell:TM:12b}).

Next, the following independent variables will be used $t=\tau_1$, $x= \zeta$ and $y=\eta$. First of all, we need to move on to real variables by putting $A_1=ae^{i\varphi}$ and $A_2=be^{i\psi}$. It results in
\begin{eqnarray}
	&& \frac{\partial a}{\partial t} + \frac{1}{2}\left[ 2\left( \frac{\partial \varphi}{\partial x} \frac{\partial a}{\partial x} -  \frac{\partial \varphi}{\partial y} \frac{\partial a}{\partial y}\right)  +a \left(  \frac{\partial^2 \varphi}{\partial x^2} -  \frac{\partial^2 \varphi}{\partial y^2}\right) \right]  =0, \label{eq:HM:Man:re:a1} \\
	&& \frac{\partial b}{\partial t} + \frac{1}{2}\left[ 2\left( \frac{\partial \psi}{\partial x} \frac{\partial b}{\partial x} -  \frac{\partial \psi}{\partial y} \frac{\partial b}{\partial y}\right)  +b \left(  \frac{\partial^2 \psi}{\partial x^2} -  \frac{\partial^2 \psi}{\partial y^2}\right) \right]  =0, \label{eq:HM:Man:re:b1} \\
	&& a \frac{\partial \varphi}{\partial t} -\frac{1}{2}\left[ \left( \frac{\partial^2 a}{\partial x^2} -  \frac{\partial^2 a}{\partial y^2}\right) - a\left( \frac{\partial \varphi}{\partial x}\frac{\partial \varphi}{\partial x} - \frac{\partial \varphi}{\partial y}\frac{\partial \varphi}{\partial y}  \right)\right] -P_1 a =0 , \label{eq:HM:Man:re:ph1}\\
	&& b \frac{\partial \psi}{\partial t} -\frac{1}{2}\left[ \left( \frac{\partial^2 b}{\partial x^2} -  \frac{\partial^2 b}{\partial y^2}\right) - b\left( \frac{\partial \psi}{\partial x}\frac{\partial \psi}{\partial x} - \frac{\partial \psi}{\partial y}\frac{\partial \psi}{\partial y}  \right)\right] -P_2 b =0,
	\label{eq:HM:Man:re:ph2}
\end{eqnarray}
where $P_{1,2} = \mu_{1,2}(a^2+b^2)$, $\mu_1=4\pi \chi^{(3)}/\varepsilon_{e}$ и $\mu_2=4\pi \chi^{(3)}/\varepsilon_{o}$

Equations (\ref{eq:HM:Maxwell:TM:12a}) and (\ref{eq:HM:Maxwell:TM:12b}), or their real form (\ref{eq:HM:Man:re:a1}) -- (\ref{eq:HM:Man:re:ph2}) have the spatially homogeneous solution:
$$ K_1= \partial\varphi/\partial x =0, \quad K_2= \partial\varphi/\partial y =0, \quad L_1= \partial\psi/\partial x =0, \quad L_2= \partial\psi/\partial y =0,$$
$$  a=a_0,\quad b=b_0, \quad \partial\varphi/\partial t = P_{10} = \mu_{1}(a_0^2+b_0^2), \quad \partial\psi/\partial t = P_{20} = \mu_{2}(a_0^2+b_0^2).$$
The usual procedure for studying stability is to linearize the initial equations and find out whether small perturbations grow or not and under what conditions this happens. In this case, the substitution is made $a=a_0+u_1$, $b=b_0+u_2$, $K_{1,2}= q_{1,2} $, $L_{1,2}=p_{1,2}$ with small values $u_{1,2}$, $q_{1,2} $, $p_{1,2} $.

Linearization of the system of equations (\ref{eq:HM:Man:re:a1}) -- (\ref{eq:HM:Man:re:ph2}) leads to the system of linear equations with respect to $u_{1,2}$, $q_{1,2}$ and $p_{1,2}$, from which the variables $q_{1,2}$ and $p_{1,2}$ can be excluded, thus obtaining only two equations for $u_{1,2}$. These standard manipulations can be omitted and only the final result can be written out:
\begin{eqnarray}
	&& \frac{\partial^2 u_1}{\partial t^2} + \frac{1}{4} \left( \frac{\partial^4 u_1}{\partial x^4} -2 \frac{\partial^4 u_1}{\partial x^2\partial y^2} + \frac{\partial^4 u_1}{\partial y^4}\right)    +\nonumber \\
	&&  \qquad ~~\qquad  +m_{11}\left( \frac{\partial^2 u_1}{\partial x^2} - \frac{\partial^2 u_1}{\partial y^2}\right) +m_{12}\left( \frac{\partial^2 u_2}{\partial x^2} - \frac{\partial^2 u_2}{\partial y^2}\right)  =0,  \label{eq:HM:Man:re:8a} \\
	&& \frac{\partial^2 u_2}{\partial t^2} + \frac{1}{4} \left( \frac{\partial^4 u_2}{\partial x^4} -2 \frac{\partial^4 u_2}{\partial x^2\partial y^2} + \frac{\partial^4 u_2}{\partial y^4}\right)   + \nonumber \\
	&& \qquad ~~\qquad  +m_{21}\left( \frac{\partial^2 u_1}{\partial x^2} - \frac{\partial^2 u_1}{\partial y^2}\right)  +m_{22}\left( \frac{\partial^2 u_2}{\partial x^2} - \frac{\partial^2 u_2}{\partial y^2}\right) =0.  \label{eq:HM:Man:re:8b}
\end{eqnarray}
where matrix elements $m_{ij}$ were introduced:
	$$m_{11} = \mu_1a_0^2,  \quad m_{12} = \mu_1a_0b_{0},  \quad m_{21} = \mu_2a_0b_0,  \quad m_{22} = \mu_2b_0^2.$$
	
Substitution expressions $u_{1,2}= B_{1,2} \exp [ik_1 x+i k_2 y -i\nu t]$  in (\ref{eq:HM:Man:re:8a}) and (\ref{eq:HM:Man:re:8b})  leads to a homogeneous system of linear equations
\begin{eqnarray}
	&&\left( \nu^2 - \frac{1}{4}\kappa^4 + m_{11}\kappa \right) B_1 + m_{12}\kappa^2 B_2 =0, \nonumber \\
	&& m_{21}\kappa^2 B_1+\left( \nu^2 - \frac{1}{4}\kappa^4 +m_{22}\kappa \right) B_2  =0, \nonumber
\end{eqnarray}
where $\kappa^2 =(k_1^2-k_2^2)$. For this system of equations to have a non-zero solution, its determinant
must be zero, which leads to the characteristic equation
	\begin{equation}\label{eq:HM:Man:re:9}
	\nu^2 = \frac{1}{4}\kappa^4  -  \frac{\kappa^2}{2}(m_{11}+m_{22})  \pm \frac{\kappa^2}{2}
	\left[ (m_{11}-m_{22})^2 +4m_{12}m_{21}\right]^{1/2} .
	\end{equation}
Given the explicit form of the matrix elements $m_{ij}$, we can obtain two dispersion relations for harmonic waves traveling at the background of a spatially homogeneous solution
$$  \nu_1^2 = \frac{1}{4}\kappa^4,\quad  \nu_2^2 = \frac{1}{4}\kappa^4 -(\mu_1a_0^2 + \mu_2b_0^2)\kappa^2.
$$	
Instability of a spatially homogeneous wave occurs under the condition
	$$ \frac{1}{4}\kappa^4 -(\mu_1a_0^2 + \mu_2b_0^2)\kappa^2 \leq  0. $$
The instability increment $G$ is defined by the expression
	\begin{equation}\label{eq:HM:Man:re:G}
	G^2= (\mu_1a_0^2 + \mu_2b_0^2)\kappa^2 - \frac{1}{4}\kappa^4.
	\end{equation}
	
The expression (\ref{eq:HM:Man:re:G}) allows us to qualitatively describe the main characteristics of MI. Since by definition $\kappa^2 =(k_1^2-k_2^2)$, the boundary of the region in the space of the components of the wave vectors $(k_1, k_2)$ in which MI takes place is determined by the equation	
$$ 
	k_1^2-k_2^2 = 4(\mu_1a_0^2 + \mu_2b_0^2).
$$
The maximum value of the increment of MI $G_m$ is given by the formula 
$$
	G^2_m = (\mu_1a_0^2 + \mu_2b_0^2)^2.
$$
The position of the maximum of the increment of MI in the plane $ (k_1, k_2) $ lies on the curve defined by the equation
$$ 
	k_1^2- k_2^2 = 2(\mu_1a_0^2 + \mu_2b_0^2).
$$	
	
For hyperbolic materials, the main values of the permittivity tensor have opposite signs. Let the case $\varepsilon_o <0$ and $\varepsilon_e >0$ be selected. In this case, the boundary of the domain in the space $(k_1, k_2)$ in which MN takes place are hyperbolas
$$
	k_1^2-k_2^2 = 16\pi \chi^{(3)}\left( \frac{a_0^2}{\varepsilon_e} - \frac{b_0^2}{\mid\varepsilon_o\mid}\right). 
$$
If we choose the case $\varepsilon_o >0$ and $\varepsilon_e <0$, then the boundary of the MI region will be hyperbolas
$$
	k_2^2-k_1^2 = 16\pi \chi^{(3)}\left( \frac{a_0^2}{\mid\varepsilon_e\mid} - \frac{b_0^2}{\varepsilon_o}\right). 
	$$
	
It is important to note that, unlike conventional (isotropic or uniaxial) media, the wave vector length of the wave that leads to MI is not limited. An exception is possible for the case of $G_m=0$, which is realized when the condition $a_0^2/b_0^2 = ~\mid \varepsilon_{e}/\varepsilon_{o}\mid$ is met:  the length of the wave vector is not limited, although there is no modulation instability.

\section{Conclusion}	
	
In this paper, the propagation of an electromagnetic wave in a nonlinear anisotropic (uniaxial) medium is considered in the case when the isofrequency surface is the hyperboloid. Such media are called hyperbolic \cite{Elser:06,Nogin:09,Drachev:13,Pod:Kivsh:13}.	If in isotropic weakly nonlinear media the propagation of weakly dispersing waves is well described by the nonlinear Schr\"{o}dinger equation, one-dimensional or two-dimensional \cite{Akh:Ank:2003,Kivsh:Agr:2005,Rys:Trub:2017}, for hyperbolic media in the same approximation, hyperbolic NLS should be used. The latter has arisen before in plasma physics and hydrodynamics. Since electromagnetic waves are vector, an adequate description of the propagation of quasi-harmonic waves is based on a system of hyperbolic NLS equations.

By the example of a simple model of a nonlinear medium, using the generalized (hyperbolic) Manakov equations, the manifestation of hyperbolicity of a dielectric material in the process of modulation instability is illustrated. 
It is shown that the geometric location of the maximum (and zero ) the values of the instability increment are hyperbolas in the plane of the wave vectors of harmonic perturbations. This means that MI occurs for the perturbation with any wave numbers in magnitude. For an ordinary media, in the one-dimensional case, such wave numbers fill a finite segment, and in the two-dimensional case, the instability region has a finite area.

In the study of MI, the hyperbolic Manakov system was considered. However, it is possible to generalize the results by redefining the matrix elements $m_{ij}$. If the contributions from the nonlinear polarization $\mathcal{P}_1$ and $\mathcal{P}_2$ are more complex functions of the amplitudes $a_0$ and $b_{0}$, then in the process of linearization of the equations (\ref{eq:HM:Man:re:a1}) -- (\ref{eq:HM:Man:re:ph2}) contributions from nonlinear polarization will generally be written as
 $ P_1 = P_{10} + P_{11}u_1 +P_{12}u_2 +\mathcal{O}(u_1^2, u_2^2)$ и $ P_2 = P_{20} + P_{21}u_1 +P_{22}u_2 +\mathcal{O}(u_1^2, u_2^2)$. The corresponding matrix elements $m_{ij}$ will be defined as
	$$m_{11} = a_0P_{11}/2,  \quad m_{12} = a_0P_{12}/2,  \quad m_{21} = b_0P_{21}/2,  \quad m_{22} = b_0P_{22}/2.$$
Then you can use(\ref{eq:HM:Man:re:9}).

Taking into account the group velocities dispersion and the more complex dependence of the polarization of the medium on the electric field strength of the wave will lead to a generalization of the equations obtained here, which can predict new phenomena in the nonlinear optics of metamaterials.

	\bigskip
\textbf{Funding}: This investigation is funded by Russian Science Foundation (project 22-11-00141).


\begin{thebibliography}{99}
	
	
	\bibitem{SSS:01} R.A. Shelby, D.R. Smith,  S. Schultz. Experimental verification of a negative index of refraction. \textit{Science}, \textbf{292}, 77 (2001).
	
	\bibitem{5} \textit{Negative-refraction Metamaterials: Fundamental Principles and Applications} Ed. by  G.V. Eleftheriades, K.G. Balmain  (N.Y.: Wiley, 2005).
	
	\bibitem{6}  V.M. Agranovich, Yu.N. Gartstein. Spatial dispersion and negative refraction of light. \textit{Phys. Usp.}  \textbf{49}, 1029 (2006).
	
	\bibitem{8}  M. Lapine,  I. V. Shadrivov, Yu. Kivshar. Nonlinear metamaterials. \textit{Rev. Mod. Phys.} \textbf{86}, 1093 (2014).
	
	
	\bibitem{Ramak:05}  S. A. Ramakrishna. Physics of negative reftactive index maaterials. \textit{Rep. Prog. Phys.} \textbf{68}, 449 (2005). 
	
	\bibitem{Sar:Shal:13}  A.K. Sarychev, Vl. M. Shalaev. \textit{Electrodynamics of metamaterials} (World Sci., New Jersey, London, Singapore, 2007). 
	
	%\bibitem{Dra:Pod:Kild:13} Drachev V.P., Podolskiy V.A., Kildishev A.V. Optics Express, 21, 15048 (2013).
	
	\bibitem{Elser:06} J. Elser, R. Wangberg, V.A. Podolskiy, E. E. Narimanov. Nanowire metamaterials with extreme optical anisotropy.  \textit{Appl. Phys. Lett.}, \textbf{89}, 261102 (2006).
	
	\bibitem{Nogin:09}  M. A. Noginov, Y. A.Barnakov, G. Zhu, T. Tumkur, H. Li, E. E. Narimanov.  Bulk photonic metamaterial with hyperbolic dispersion . \textit{Appl. Phys. Lett}. \textbf{94} 151105 (2009).
	
	\bibitem{Drachev:13} V. P. Drachev, V. A. Podolskiy,  A. V.  Kildishev. Hyperbolic metamaterials: new physics behind a classical problem. 	\textit{Opt. Express}, \textbf{21}, 190862 (2013)
	
	\bibitem{Pod:Kivsh:13}  A. Poddubny, I. Iorsh, P. Belov, Yu.Kivshar. Hyperbolic metamaterials. 
	\textit{Nat. Photon.} \textbf{7}, 958 (2013)
	
	\bibitem{Pod:Belo:12} A. N. Poddubny, P. A. Belov, P. Ginzburg, et al. 	Microscopic model of Purcell enhancement in hyperbolic metamaterials, \textit{Phys. Rev. B. } \textbf{86}, 035148 (2012).
	
	\bibitem{Pod:Belo:13} A. Poddubny, P.V.  Belov, Yu. S. Kivshar. Purcell effect in wire metamaterials. 
	\textit{Phys. Rev. A.} \textbf{87}, 035136 (2013).
	
	\bibitem{Zhao:2013}  J. Zhao, H. Zhang, X. Zhang, D. Li, H. Lu, and M. Xu, Abnormal behaviors of Goos-Hänchen shift in hyperbolic metamaterials made metamaterials. \textit{Opt. Express} \textbf{21}, 19113 (2013).
	
	\bibitem{Ishii:14}  S. Ishii, M.Y. Shalaginov, V. E. Babicheva, A. Boltasseva, and A. V.Kildishev, Plasmonic waveguides cladded by hyperbolic metamaterials. \textit{Opt. Lett.} \textbf{39}, 4663 (2014).
	
	\bibitem{Babicheva:15}   V. E. Babicheva,  M. Y. Shalaginov,  S. Ishii, A. Boltasseva, and A. V.Kildishev, Finite-width plasmonic waveguides with hyperbolic multilayer cladding. \textit{Opt. Express} \textbf{23}, 9681 (2015).
	
	\bibitem{Boardm:17} A. D. Boardman,  A. Alberucci,  G. Assanto, V V Grimalsky, B Kibler, J McNiff, I S Nefedov, Yu G Rapoport and C A Valagiannopoulos. Waves in hyperbolic and double negative metamaterials including rogues and solitons. 
	\textit{Nanotechnology} \textbf{28}, 444001 (2017).
	
	\bibitem{Lyash:Maim:15} E.I. Lyashko, A.I. Maimistov. Linear guided waves in a hyperbolic slab waveguide. Dispersion relations. 	\textit{Quantum Electronics} \textbf{45}, 1050 (2015).
	
	\bibitem{Lyash:Maim:16} E.I. Lyashko, A.I. Maimistov. Guided waves in asymmetric hyperbolic slab waveguides. The TM mode case. 	\textit{J. Opt. Soc. Am. B.} \textbf{33}, 2320 (2016). 
	
	\bibitem{Lyash:Maim:17} E.I. Lyashko, A.I. Maimistov. Modes of a nonlinear planar waveguide with a dielectric layer immersed in a hyperbolic medium. \textit{Quantum Electronics} \textbf{47} 1053 (2017). 
			
	\bibitem{Tekin:21}	Bayram Tekin. Hyperbolic metamaterials and massive Klein-Gordon equation in (2 + 1)-dimensional de Sitter spacetime. \textit{Phys.Rev. D.} \textbf{104}, 105004 (2021)
	
	\bibitem{Tamir:78} T. Tamir (Ed.).  \textit{Integrated Optics} (Berlin: Springer, 1983; Moscow: Mir, 1978)
	
	\bibitem{Sukhorukov:Vinog:90} M.B. Vinogradov, O.V. Rudenko, A.P. Sukhorukov. \textit{Wave theory} ( Moskov. Nauka, 1990 ) .
	
	\bibitem{Maim:Bashar:book}  A. I. Maimistov, A.M. Basharov. \textit{Nonlinear Optical Waves} ( Kluwer Academic Publishers, Dortrecht, Boston, London, 1999).
	
	\bibitem{Akh:Ank:2003}  N.N. Akhmediev, A. Ankiewicz. \textit{Solitons. Nonlinear pulses and beams} ( Chapman \& Hall, London, Weinheim, New York, Tokyo, Melbourne, Madras, 1997).
	
	\bibitem{Kivsh:Agr:2005} Yu. S. Kivshar, G.P. Agrawal. \textit{Optical solitons. From Fiberrs to Photonic Crystals} ( Academic Press. 2003) .
	
	\bibitem{Rys:Trub:2017} N. M. Ryskin, D. I. Trubetskov. \textit{Nonlinear waves} (URSS. LENAND, Moscow, 2017).
	
	
	\bibitem{Boardman:2016} A. Alberucci, Ch. P. Jisha, A.D. Boardman, and G. Assanto. Anomalous diffraction in hyperbolic materials. \textit{Phys. Rev. A} \textbf{94}, 033830 (2016).
	
%	\bibitem{Boardman:2017} A D Boardman, A Alberucci, G Assanto, V V Grimalsky, B Kibler,
%	J McNiff, I S Nefedov, Yu G Rapoport and C A Valagiannopoulos.  Waves in hyperbolic and double negative
%	metamaterials including rogues and solitons.  \textit{Nanotechnology}, \textbf{28},  444001 (2017).
		

	
	% Симуляции Гравитации и  голографического принципа
	
	\bibitem{Smol:Narim:10} I.I. Smolyaninov, E.E. Narimanov.  Metric signature transitions in optical metamaterials.  
	\textit{Phys. Rev. Lett.} \textbf{105}, 067402 (2010).
	
	\bibitem{Smol:Hun:11}  I.I. Smolyaninov, Y.J. Hung.  Modeling of time with metamaterials. 
	\textit{J. Opt. Soc. Am. B.} \textbf{28}, 1591 (2011).
	
	\bibitem{Smol:11} I.I. Smolyaninov. Vacuum in a Strong Magnetic Field as a Hyperbolic Metamaterial.  
	\textit{Phys. Rev. Lett.} \textbf{107}, 253903 (2011).
	
	\bibitem{Smol:Hwan:12}  I.I. Smolyaninov, E. Hwang, E.E. Narimanov.   Hyperbolic metamaterial interfaces: Hawking radiation from Rindler horizons and spacetime signature transitions.  \textit{Phys. Rev. B.} \textbf{85}, 235122 (2012).
	
	\bibitem{Smol:12}  I.I. Smolyaninov.  ‘Planck-scale physics’ of vacuum in a strong magnetic field.  
	\textit{Phys. Rev. D.} \textbf{85}, 114013 (2012).
	
	\bibitem{Smol:Hun:13} I.I. Smolyaninov, Y.J. Hung. 	Minkowski domain walls in hyperbolic metamaterials 
	\textit{Phys. Lett. A.} \textbf{377}, 353 (2013).
	
	\bibitem{Smol:Yost:13}	I.I. Smolyaninov, B. Yost, E. Bates, V.N. Smolyaninova.  Experimental demonstration of metamaterial “multiverse”\, in a ferrofluid.  \textit{Opt. Express} \textbf{21}, 14918 (2013).
	
	\bibitem{Smol:13b} I.I. Smolyaninov.  Modeling of causality with metamaterials. \textit{J. Opt.} \textbf{15}, 025101 (2013)
	
	\bibitem{Smol:13} I.I. Smolyaninov. Analog of gravitational force in hyperbolic metamaterials. 	\textit{Phys.Rev. A.} \textbf{ 88}, 033843 (2013)
	
	\bibitem{Smol:16} I.I. Smolyaninov.	Holographic duality in nonlinear hyperbolic metamaterials, 	\textit{J. Opt.} \textbf{16}, 075101 (2014)
	
	
	% Гиперболическое НУШ
	
	\bibitem{AiLin:2010} G. Ai-Lin ,  L. Ji.  Exact solutions of (2+1)-dimensional HNLS equation, 
	\textit{Commun. Theor. Phys.} \textbf{54}, 401 (2010).
	
	\bibitem{Ablowitz:Ma:2016}	 M. J. Ablowitz, Yi-Ping Ma, I. Rumanov. \textit{A universal asymptotic regime in the hyperbolic nonlinear Schrodinger equation. arXiv:1606.02782 [nlin.PS].}
	
	\bibitem{Aliuy:2018}  A.I. Aliyu, M. Inc, A. Yusuf, D. Baleanu. Optical solitary waves and conservation laws to the (2+1)-dimensional hyperbolic nonlinear Schr\"{o}dinger equation, 
	\textit{Mod. Phys. Lett. B.} \textbf{32}, 1850373 (2018).
	
	\bibitem{Apeanti:2019}.  W. O. Apeanti,  A. R. Seadawy, D. Lu.
	 Complex optical solutions and modulation instability of hyperbolic Schr\"{o}dinger dynamical equation, 
	\textit{Results Phys.} \textbf{12}, 2091 (2019).
	
	\bibitem{Durur:2020} H. Durur, E. Ilhan, H. Bulut. 
	 Novel complex wave solutions of the (2+1)-dimensional hyperbolic nonlinear Schr\"{o}dinger equation,
	\textit{Fractal Fract.} \textbf{4}, 41 (2020).
	
	\bibitem{TalaTebue:2020} E. Tala-Tebue, C. Tetchoka-Manemo, H. Rezazadeh, A. Bekir, Y.M. Chu. 
	 Optical solutions of the (2+1)-dimensional hyperbolic nonlinear Schr\"{o}dinger equation using two different methods, 
	\textit{Results Phys.} \textbf{19}, 103514 (2020).
	
	%\bibitem{Rehman:2020} H. Ur Rehman, M. A. Imran, N. Ullah, A. Akg\"{u}, Exact solutions of (2+1)-dimensional %Schr\"{o}dinger’s hyperbolic equation using different techniques, Numer. Meth. Part. Differ. Equ., 2020
	
	
	\bibitem{Baleanu:Park:2021}  D. Baleanu,  K. Hosseini,  S. Salahshour, Khadijeh Sadri, Mohammad Mirzazadeh, Choonkil Park, and Ali Ahmadian. The (2+1)-dimensional hyperbolic nonlinear Schr\"{o}dinger equation and its optical solitons, 
	\textit{AIMS Mathematics}, \textbf{6}, 9568 (2021). 
	
	\bibitem{Manakov:73} S. V. Manakov, On the theory of two-dimensional stationary
	self-focusing of electromagnetic waves, Zh. Eksp. Teor. Fiz.\textbf{65}, 505--516 (1973) [Sov.Phys. JETP \textbf{38}, 248--253 (1974)].
	
	
	
\end{thebibliography}
\end{document}